\documentclass[twocolumn]{aastex701}
\usepackage{apjfonts}
\usepackage{rotating}
\usepackage{amsmath, color,graphicx,tablefootnote,subfigure}
\usepackage{graphics,subfigure,hyperref,float}
\usepackage[rightcaption]{sidecap}
\usepackage{tabularx}

\newcommand{\CH}[1]{\colhead{#1}}
\newcommand{\D}{$^{\dagger}$}

\newcommand{\W}{$\lambda$}
\newcommand{\HII}{H$\,${\sc ii}}

\newcommand\yp{$Y_{\rm p}$}
\newcommand\te{T$_e$}
\newcommand\den{n$_ e$}

\def\pb{{P$\beta$}}
\def\10830{{He~I $\lambda$10830}}
\def\pg{{P$\gamma$}}
\def\pd{{P$\delta$}}
\def\y+{\ensuremath{\mathrm{y}^{+}}}

\shorttitle{The LBT \yp\ Project III}
\shortauthors{Weller et al.}

\begin{document}

\title{The LBT \yp\ Project III: LUCI Spectra of Metal-Poor Nebulae}

%
%

%
\author[0000-0003-4912-5157]{Miqaela K.\ Weller}
\affiliation{Department of Astronomy, The Ohio State University, 140 W 18th Ave., Columbus, OH, 43210}
\email{weller.133@buckeyemail.osu.edu}
\author[0000-0003-1435-3053]{Richard W.\ Pogge}
\affiliation{Department of Astronomy, The Ohio State University, 140 W 18th Ave., Columbus, OH, 43210}
\affiliation{Center for Cosmology \& AstroParticle Physics, The Ohio State University, 191 West Woodruff Avenue, Columbus, OH 43210}
\email{pogge.1@osu.edu}
\author[0000-0003-0605-8732]{Evan D.\ Skillman}
\affiliation{Minnesota Institute for Astrophysics, University of Minnesota, 116 Church St. SE, Minneapolis, MN 55455}
\email{skill001@umn.edu}
\author[0009-0006-2077-2552]{Erik Aver}
\affiliation{Department of Physics, Gonzaga University, 502 E Boone Ave., Spokane, WA, 99258}
\email{aver@gonzaga.edu}
\author[0000-0002-0361-8223]{Noah S.\ J.\ Rogers}
\affiliation{Center for Interdisciplinary Exploration and Research in Astrophysics (CIERA), Northwestern University, 1800 Sherman Avenue, Evanston, IL 60201, USA}
\email{noah.rogers@northwestern.edu}

\author[0000-0002-4153-053X]{Danielle A. Berg}
\affiliation{Department of Astronomy, The University of Texas at Austin, 2515 Speedway, Stop C1400, Austin, TX 78712, USA}
\email{daberg@austin.utexas.edu}
\author[0000-0001-8483-603X]{John J. Salzer}
\affiliation{Department of Astronomy, Indiana University, 727 East Third Street, Bloomington, IN 47405, USA}
\email{josalzer@iu.edu}
%
%
\author[0000-0002-2901-5260]{John H. Miller, Jr.}
\affiliation{Minnesota Institute for Astrophysics, University of Minnesota, 116 Church St. SE, Minneapolis, MN 55455}
\email{mill9614@umn.edu}
\author{Jayde Spiegel}
\affiliation{Department of Astronomy, The Ohio State University, 140 W 18th Ave., Columbus, OH, 43210}
\email{spiegel.85@buckeyemail.osu.edu}
%


\begin{abstract}
Accurately determining the elemental abundances of a low metallicity nebula strongly depends on measuring the density (\den) and temperature (\te) of the gas. Because these two parameters are inherently degenerate when derived solely from H and He recombination lines, we rely on the density-sensitive \10830\ line to assist in resolving this issue, especially for accurate He abundances. To facilitate this, we present near-IR (NIR) LUCI spectra of 48 low-metallicity targets from the Large Binocular Telescope (LBT) and homogeneously reduce them using Pypeit as part of the LBT \yp\ Project. IR spectra require special care, and we wavelength calibrate by-hand using the bright OH emission lines, carefully apply proper telluric corrections, and co-add the spectra of LUCI1 and LUCI2 on a resampled grid to ensure accurate results. We use a Gaussian profile to fit the emission lines and measure the fluxes relative to Paschen-gamma (\pg), resulting in \10830\ to \pg\ ratios consistent with previous studies. As a result, this work significantly expands the available dataset of NIR \10830\ fluxes in low metallicity galaxies. These high-quality measurements, where we find a median flux ratio uncertainty of $\tilde{\sigma} = 0.08$, reduce the overall uncertainties in helium abundance estimates for individual targets. The increased size of the high-quality sample enables searching for systematic uncertainties and improves the reliability of the helium abundance determinations used to infer the primordial helium abundance (\yp).
\end{abstract}



\section{Introduction}\label{sec1}

Measuring the primordial helium abundance, \yp, using low-metallicity galaxies is an important cross-check for the standard cosmological model. This methodology follows from \citet{Peimbert1974, Peimbert1976}, where helium abundances are plotted against a metal, like oxygen, to extrapolate to zero metallicity in order to determine \yp. This requires accurate abundances that depend on the temperature and density of the nebular gas. \citet{aver2010} demonstrated that there is a fundamental degeneracy between temperature and density that limits the precision of this approach. \citet{aver2011} introduced the use of a Markov Chain Monte Carlo methodology to appropriately estimate uncertainties accounting for such degeneracies.
The goal then becomes to constrain temperature and density as well as possible to minimize uncertainties on the derived helium abundance.

\citet{izot2014} demonstrated that observations of the NIR \10830 emission line led to stronger constraints on the derived density.
Because emission in \10830 is enhanced through collisional excitation of electrons in the meta-stable 2 $^3$S level, the emissivity of the \10830 line is very sensitive to density. Therefore, \10830 places strong
constraints on gas density and resolves the temperature-density degeneracy of the
other He I lines.  
\citet{aver2015} explored the impact of using this powerful constraint in producing a new value of \yp.

In principle, adding a NIR emission line to optically observed emission lines for the purpose of deriving abundances can be quite challenging due to atmospheric transparency variations in the NIR and matching of apertures.  
These concerns are greatly reduced because of the spectral proximity of the \ion{H}{1} Paschen $\gamma$ (hereafter \pg) emission line at $\lambda$ 10941\AA to \10830. Because these two lines are so close together in wavelength, their reddening-insensitive ratio can be measured very accurately and can then be tied to the abundance analysis through the theoretical ratios of the Case B \ion{H}{1} recombination lines.

Thus, the goal for the LBT \yp\ project is to obtain high quality \10830 observations that have complementary optical spectra from the LBT's Multi-Object Double Spectrograph (MODS). The LBT LUCI (LBT 
Utility Camera in the Infrared) spectrographs are well-suited for this purpose and have already been applied to Leo~P \citep{aver2021}, where the inclusion of \10830 reduced the uncertainty on the helium abundance mass fraction, Y, by 70\%. To date, we have obtained a larger sample of 48 new LUCI spectra, covering most of our 60 MODS spectra. For those that we do not acquire with LUCI, we use literature \10830 values, as was done with AGC 198691 \citep{aver2022}, where near-IR fluxes were collected by \citet{Hsyu2020}.

The format of this paper is organized as follows. In \S\ref{reduction} we describe the LUCI data reduction and analysis, including sky subtraction, object extraction, and telluric correction. In \S\ref{ratios}, we measure the line ratios of \10830\ to \pg\ that will be used for the He abundance analysis. We compare our measured ratios from the LBT \yp\ Project to the literature in \S\ref{compare}, ultimately providing the most robust catalog of near-IR fluxes for low-metallicity galaxies.

\section{Data Reduction and Analysis}

Paper~I \citep{Skillman2026} details the overall target selection and observational procedures for the LBT-LUCI spectra of the 48 galaxies used in this analysis. A brief summary is provided below.

Over the course of our program, we gather near-IR data on 48 \HII{} regions in metal-poor galaxies to primarily measure the \10830 and \pg\ lines. We aimed to obtain a total integration time of one hour on each of the LUCI spectrographs for each of the galaxies presented. This was achieved through six 600-second exposures, taken with offsets ranging from 5" to 20". These offsets were performed to achieve a dither pattern of ABBAAB to mitigate night-sky lines. All data were gathered using the N1.8 camera (0.25" per pixel) with the 1" long-slit mask and the zJspec filter, covering a wavelength range of 0.95 $\micron$ to 1.35 $\micron$ in which the dispersion is 2.323 {\AA} per pixel. 

For calibration frames, we have two sets: closed-dome and on-sky. The closed-dome calibrations consist of darks, flats, and arcs, while on-sky calibrations consist of telluric standard stars at similar airmass to each target, selected using the Gemini Observatory Telluric Standard Search tool\footnote{\url{https://www.gemini.edu/observing/resources/near-ir-resources/spectroscopy/telluric-standard-search}}. In 10 cases, tellurics were shared across a sequence of galaxies that were close on the sky for efficiency in observations. Because we have sufficient resolution, we can also achieve wavelength calibration using the bright OH emission lines alone. This process was done by hand, identifying 5-6 bright OH lines with an overall RMS of about 0.18 pixels. We routinely examined the arc frames, but we found that the wavelength calibration we achieve using the instantaneous night skylines is equally effective, if not better, than the arcs that were taken at zenith, without flexure. In fact, after the first few iterations of our analysis, we stopped using the arc frames altogether. We continued to acquire these calibration frames to ensure the spectrographs were configured correctly. 

\subsection{Data Reduction}
\label{reduction}

For reductions, we utilize {\tt PypeIt}\footnote{GitHub: \url{https://github.com/pypeit/PypeIt}}, a semi-automated Python package for slit-based spectroscopy \citep{pypeit:joss_arXiv, pypeit:zenodo, pypeit:joss_pub}. {\tt PypeIt} supports a suite of instruments that includes both LUCI and  MODS on the LBT, but we only use {\tt PypeIt} for the reduction of the LUCI data. {\tt PypeIt} provides instrument-specific default configurations for both LUCI1\footnote{LUCI1 Configuration: \url{https://pypeit.readthedocs.io/en/release/pypeit_par.html\#lbt-luci1-lbt-luci1}} and LUCI2\footnote{LUCI2 Configuration: \url{https://pypeit.readthedocs.io/en/release/pypeit_par.html\#lbt-luci2-lbt-luci2}}. We start with the same adjustments to these configurations for each target. First, we configured {\tt PypeIt} to perform image-differencing based on the offsets in our observations. This allows for the accurate subtraction of bright night-sky lines where {\tt PypeIt} fits the sky-residuals. The objects are masked in this process, so our sky regions consisted of pixels not containing an object. We adjust the break-point spacing of the B-Spline fits to optimize the quality of the sky-subtraction. Second, since we wavelength calibrated by hand, we ensured {\tt PypeIt} used our fit in its algorithm. We enable the identification of cosmic rays as well. 18/48 of the spectra acquired were bright, isolated, high S/N compact objects that easily fit on the slit and therefore required no modifications to these described extraction parameters within {\tt PypeIt}. In these targets, the automated object-finding algorithm succeeded and was able to mask the objects for sky-subtraction.

\begin{figure*}
  \centering
  \includegraphics[width=0.7\textwidth]{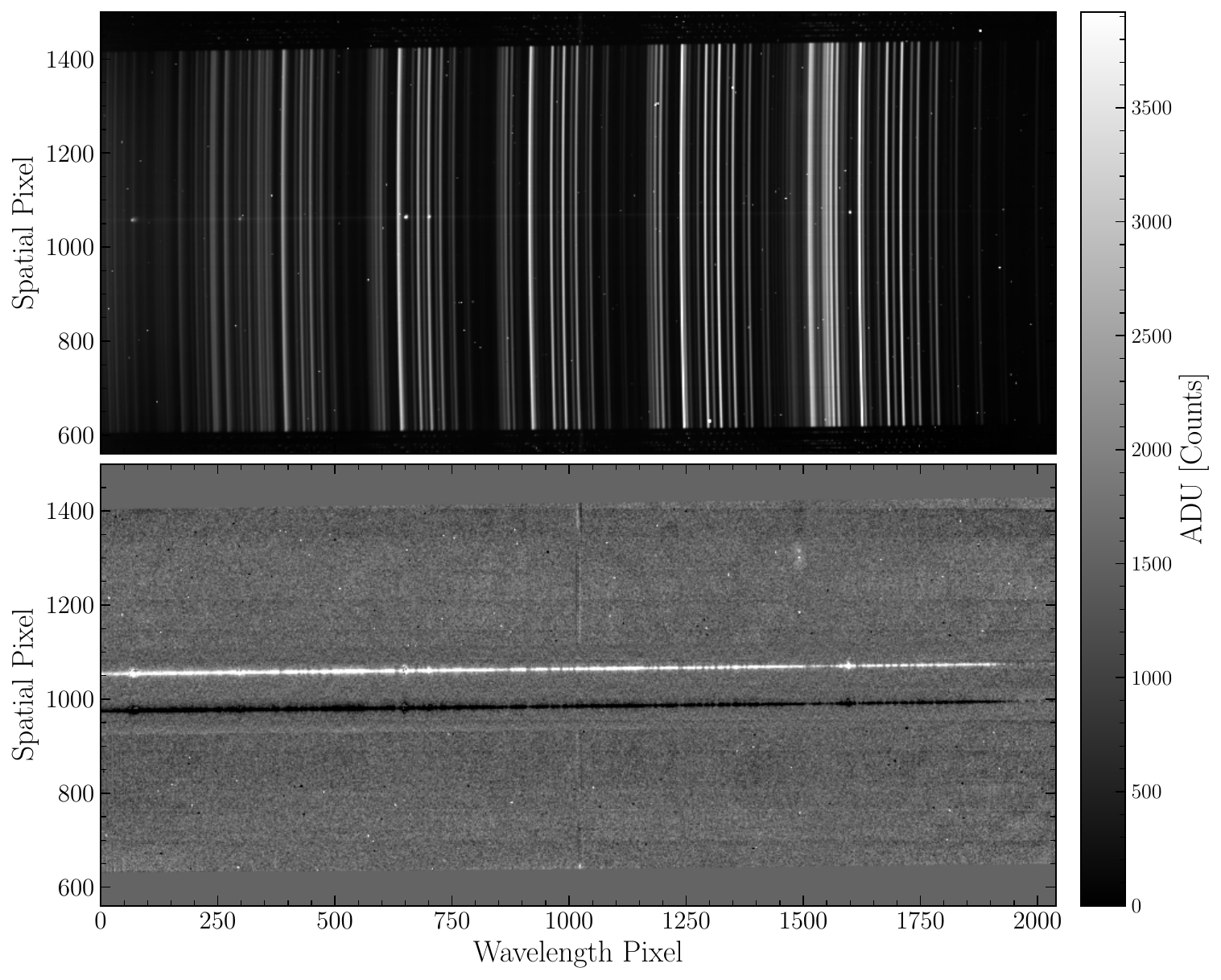}\\
  (a) \textit{Top:} A single, raw 2D spectrum from LUCI1 for UM461. \textit{Bottom:} The sky-subtracted 2D spectrum for the same frame described previously. The object trace is visible in both images. \\[1em] 
  \includegraphics[width=0.7\textwidth]{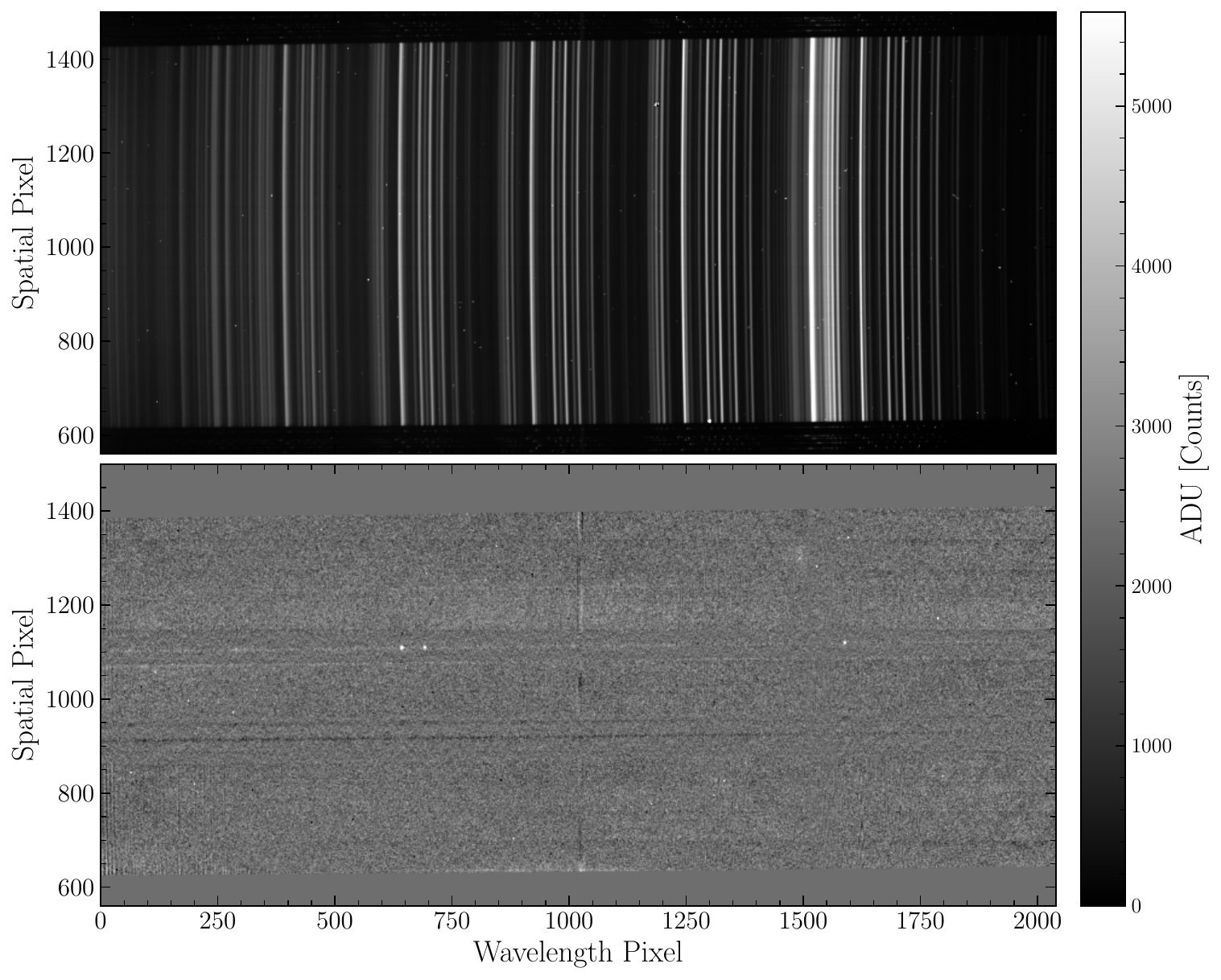}\\
  (b) \textit{Top:} A single, raw 2D spectrum from LUCI1 for DDO68. \textit{Bottom:} The sky-subtracted 2D spectrum for the same frame described previously. This object required manual extraction as the trace was too faint to be identified by the automated algorithm. 
  \caption{The bright, night-sky OH lines seen in the top panels are removed when sky-subtraction is performed, showing the success of the algorithm. Note that in the IR, {\tt Pypeit} performs sky-subtraction by fitting the residuals produced by image-differencing. This is the cause of the black traces.}
  \label{fig:raw_spectra}
\end{figure*}

30/48 galaxies required changes to our standard \texttt{PypeIt} pipeline. However, 20 of these 30 targets required the slit edges to be redefined after being improperly identified and/or required manual extraction, despite these targets being point sources. It is important to properly define the slit edges for optimal sky-subtraction. When this needed to be done, we noted the x-value of our preferred left and right edges and modified the input file accordingly. Finally, manual extraction was required if the object was too faint to be identified by the automated algorithm in \texttt{PypeIt}. In all cases, the emission lines were easily visible by eye. Therefore, we positioned the trace to intersect the center of the lines and picked the appropriate boxcar radius based on the width of the emission lines. The other 10 objects had extended regions or other objects on the slit, requiring much more intervention by hand. In these cases, we had to tailor the reductions individually and conditioned the runs for each galactic target. Specific targets and modifications are listed below.

\textit{LEDA101527.} We obtained the spectra for this target on May 1, 2022. The LBT was closed for most of the night due to high winds. Before closing, we were able to acquire 4 out of 6 exposures, making the total integration time only 40 minutes on each of the two spectrographs. We still achieve an ABBA dither pattern and modify the input file to reflect this. Despite the lower exposure time, our signal was strong enough to measure the emission lines.

\textit{Mrk71.} We attempted to gather this target on October 14, 2022. During acquisition, the spectra appeared to be saturating the detector. We chose to re-acquire this object on November 14, 2022 with an exposure time of 300s instead of 600s, making the total integration time only 30 minutes. 

\textit{HS2236.} During reduction, the automated algorithm identified one object with emission lines. In inspecting the 2D-spectrum, we noticed another object very nearby with the same emission lines. We chose to manually extract this second object in order to mask the region surrounding this object for better sky-subtraction. We follow the same post-processing steps for each object, including measuring the line ratio. The object we include in our analysis is chosen from the detailed targeting information from the observations themselves.

\textit{KUG0743.} Similar to HS2236, there were two targets on the slit. However, manual extraction was needed as the automated algorithm was not able to identify both objects without intervention. Again, both had the line ratios measured, but the object chosen for analysis is the one based on targeting details.

\textit{SBS1159.} This object had an extra LUCI2 observation. Because we perform image differencing, we located the position of the trace on the slit to determine if any dither pattern was followed. We found that an extra A frame was acquired, making the dither pattern AABBAAB instead of ABBAAB. For this specific frame, we use the first B exposure for the background.

\textit{IZw18.} Due to no other suitable targets being available during the night we gathered this object, we got an extra 3x10 minute exposures. However, 2/9 frames had to be thrown out due to an artifact in acquisition that could not be removed through processing, leaving us with 7x10 minute frames. This object also had two objects on the slit and we once again extract both but report the south-east object in our analysis.

\textit{WISEAJ085115 and Mrk5.} These two targets had a star on the slit. We allowed {\tt PypeIt} to extract both the star and our object for proper sky-subtraction. The star is not used in the subsequent post-processing steps.

\textit{SDSSJ2104.} The observers on November 2, 2024 had extra time before the next target, so we acquired 10x10 minute exposures for this target with an ABBA sequence followed by our usual ABBAAB pattern.

\textit{DDO68.} After a series of command errors at the telescope on March 19, 2022, we gathered 8x10 minute exposures on LUCI1 and 7x10 minute exposures on LUCI2. There was a mistake in the offsets, so for LUCI1, we end up with a dither pattern of ABAAAABC. For LUCI2, the dither pattern is BAAAABC. For the C frame, we use the nearest exposure in time for the background, making this the B frame. 


\textit{SBS1331.} This target had a cosmic ray in the middle of the \10830 emission line. \texttt{PypeIt} generally uses the \textsc{lacosmic} routine \citep{vanD2001} with extra machinery to handle false positives. For this specific case in SBS1331, we simply do not allow the cosmic ray to expand in the normal routine. This does not change the flux of the line and therefore does not affect subsequent post-processing steps.

In Figure \ref{fig:raw_spectra}, we illustrate the success of our sky-subtraction and extraction parameters by first showing a raw 2D spectrum contrasted with the sky-subtracted science frame for a case where manual intervention was not needed (top, UM461) and a case where extensive manual intervention was required (bottom, DDO68). Both raw 2D spectra show a plethora of night-sky OH lines. For UM461, the object trace is clearly visible in the raw image and is easily identified by the automated algorithm, but this is not true for the much fainter DD068. After performing sky-subtraction through image differencing, which results in the appearance of black traces, all night-sky lines are effectively removed. In the case of DDO68, this enables the bright emission lines of our target to finally become visible, allowing us to manually extract the object. We chose a boxcar extraction window based on the size of the emission lines to create our 1D spectra.

\subsection{Flux Calibration \& Telluric Correction}

\begin{figure}
\centering
\centerline{\includegraphics[width=0.9\columnwidth]{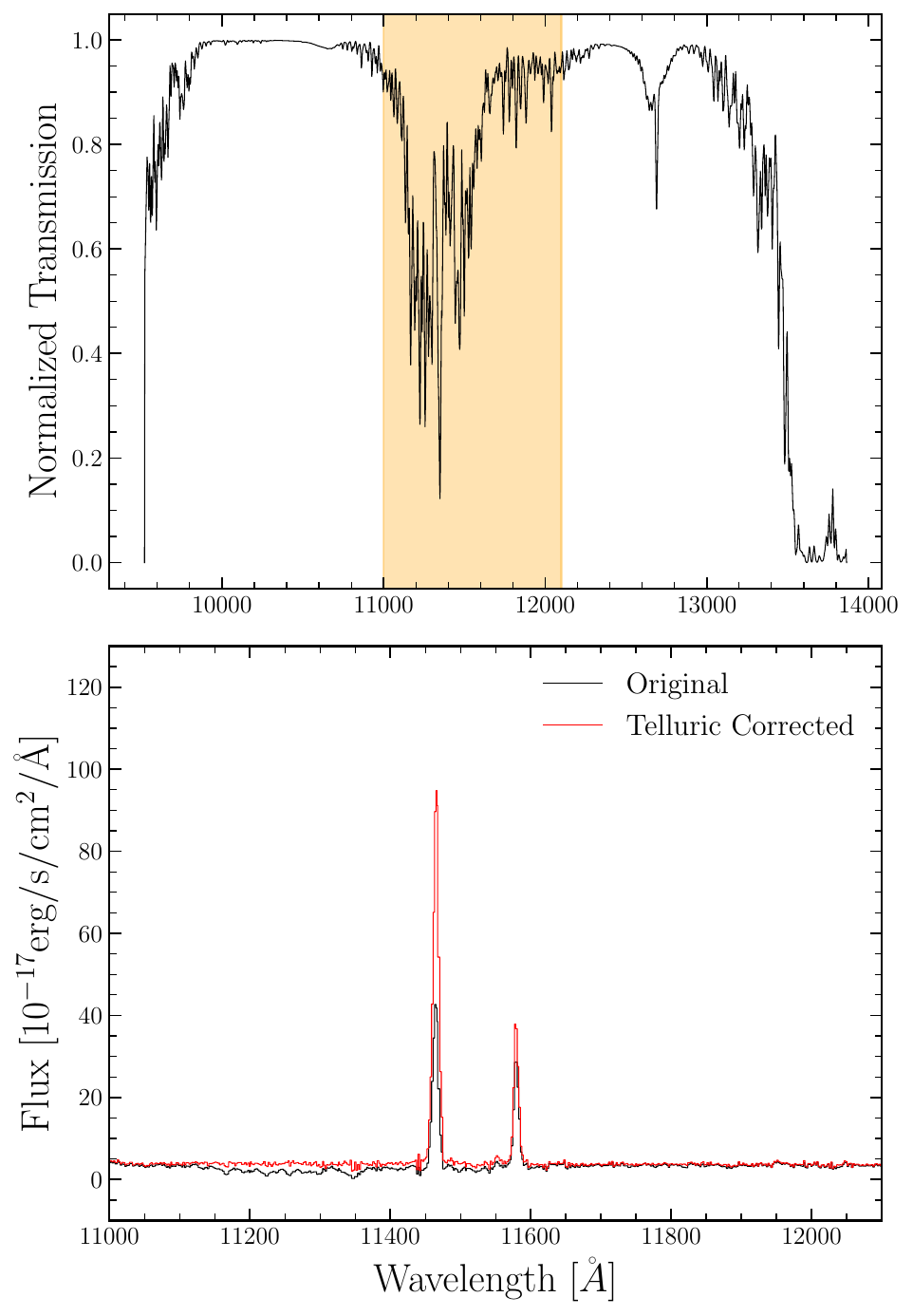}}
    \caption{\textit{Top:} Telluric model generated from {\tt PypeIt}. \textit{Bottom:} Comparison of our original un-corrected spectrum for UM420 ($z \sim 0.058$) in black with our corrected spectrum in red.}
    \label{fig: telluric_ex}
\end{figure}

Our main objective is to measure the flux ratio of \10830 to the nearby \pg\ line. This eliminates the need for a perfectly robust end-to-end flux calibration, unlike with LBT-MODS. However, we still perform a flux calibration in order to report the ratio of \pb\ and \pd\ to \pg\ that span our entire wavelength range. To do this, we generate a sensitivity function using the known flux and counts from the associated telluric star and apply it to our science spectra.

\begin{figure*}
\begin{tabularx}{1\textwidth}{cc}
\centering
  \includegraphics[width=85mm]{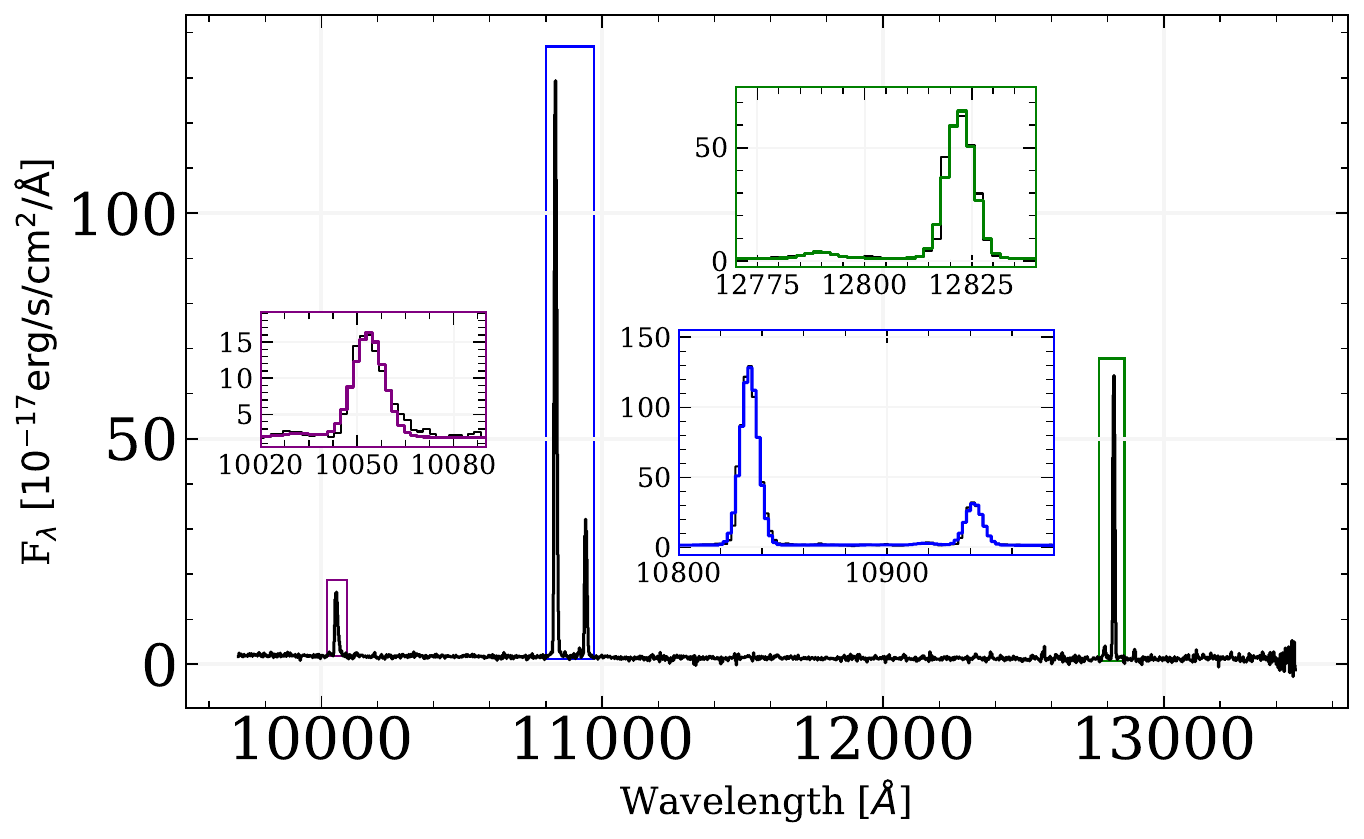} &   \includegraphics[width=85mm]{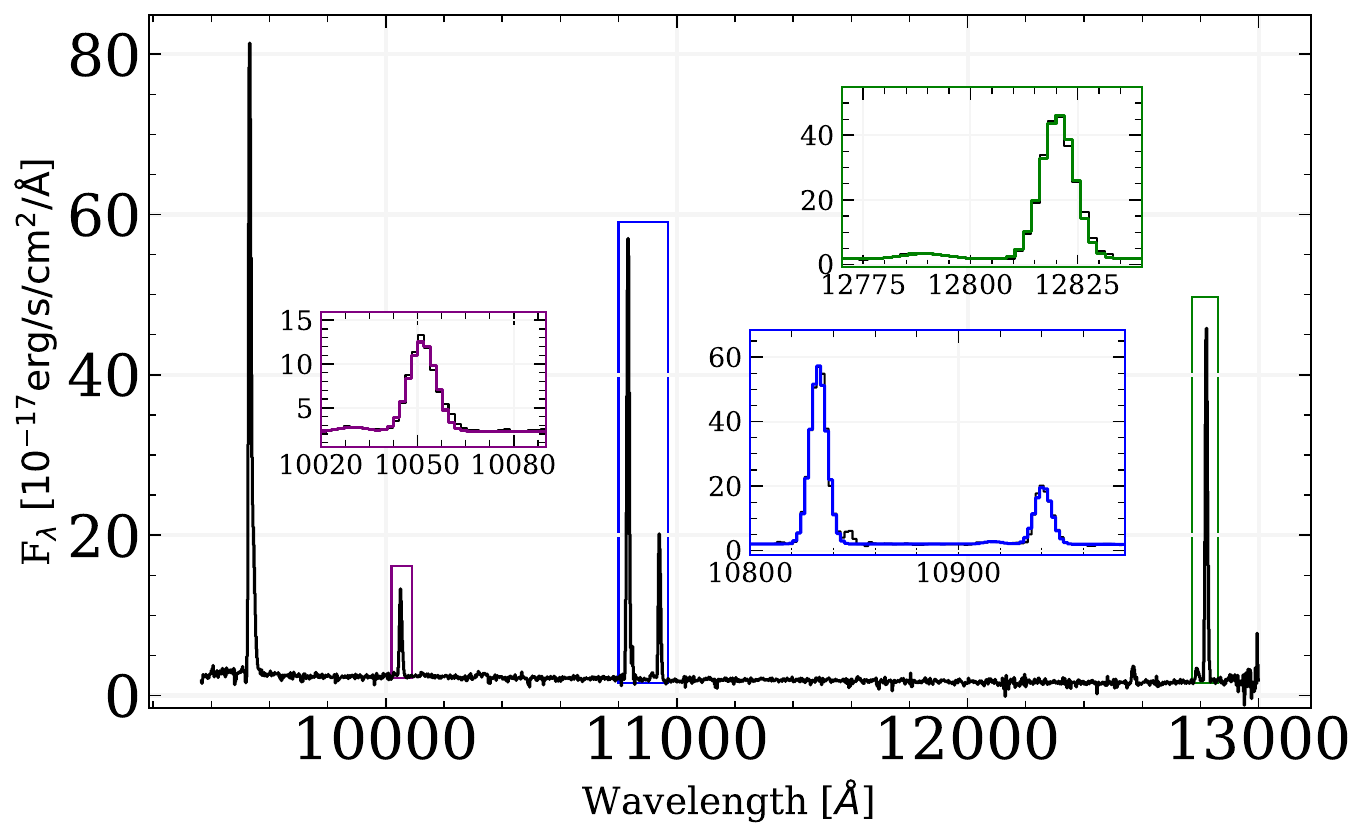} \\
(a) Final spectrum for HS 0122+0743 from LUCI1 & (b) Final spectrum for SBS 0948+532 from LUCI2 \\[6pt]
\multicolumn{2}{c}{\includegraphics[width=85mm]{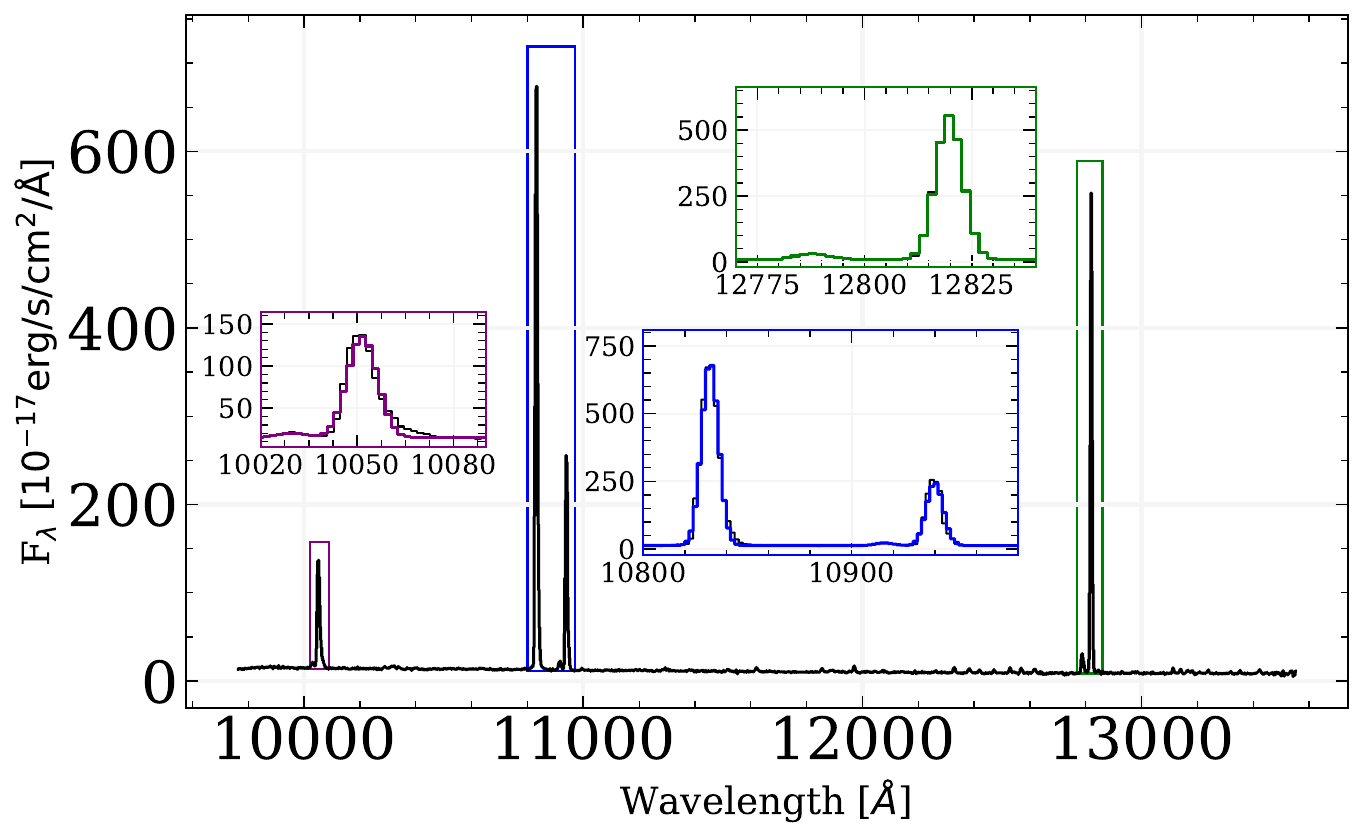} }\\
\multicolumn{2}{c}{(c) Final spectrum for UM461 from LUCI1+2}
\end{tabularx}
\caption{Examples of our final, post-processed spectra from LUCI1, LUCI2, and LUCI1+2 showing our fits to the line profiles.}
\label{fig:spectra}
\end{figure*}

With our flux-calibrated spectra, we co-add the boxcar extracted spectra to increase our signal. After this, the last step in our post-processing is to apply a telluric correction. This correction is absolutely critical for our ground-based near-IR data since the absorption bands caused by Earth's atmosphere (e.g., H$_2$O and O$_2$) are strong in this wavelength range. This is especially true for our higher-redshift targets (e.g., $z \gtrsim 0.016$). {\tt PypeIt} uses model-based telluric correction, which means that an observatory-specific telluric grid (i.e., Mount Graham from 9000\textup{~\AA} to 25,000\textup{~\AA}) and user-defined model are used. The telluric grids can be downloaded via {\tt PypeIt}. In Figure \ref{fig: telluric_ex}, we show the telluric model chosen by {\tt PypeIt} in the top panel and the telluric-corrected spectrum for UM420 in the bottom panel. UM420 has a redshift of $z \sim 0.058$, which places \10830 in a region of strong telluric contamination. However, not only is the continuum flatter compared to our un-corrected spectrum, but the flux of \10830\ and \pg\ are clearly recovered after being contaminated with telluric absorption. Indeed, the measured ratio increases from 1.66 to 2.8, a factor of $\sim 1.7$, underscoring the need for telluric correction.

\subsection{Co-addition of LUCI1 and LUCI2}\label{coadd}

33 of our 48 targets were acquired with both LUCI1 and LUCI2. However, throughout our years of observations, one of LUCI1 or LUCI2 were periodically off-telescope for repair or maintenance. For our sample, 9 out of our 48 galaxies were only observed using LUCI1, while 6 of our 48 targets were only observed using LUCI2. For these monocular observations, post-processing ended before this co-addition step.

For the 33 galaxies where we have LUCI1 and LUCI2 spectra, we combine the 1D, flux-calibrated, and telluric-corrected spectra from both instruments to improve the overall signal quality. These spectrographs have slightly different dichroic cut-ons (0.89$\mu m$ for LUCI1 and 0.95$\mu m$ for LUCI2) and differences in their wavelengths due to slight flexure shifts. We therefore had to resample the spectra onto a uniform grid from 0.98 $\micron$ to 1.36 $\micron$ before the co-addition to reconcile these differences. We used the {\tt FluxConservingResampler} module from {\tt specutils} that redistributes the flux according to the fractional overlap between the input and preferred wavelength grids such that the total integrated flux is conserved.

\section{Line Ratios}
\label{ratios}

To measure the line fluxes, a Gaussian profile was used 
\begin{equation}
    F_\lambda = \mbox{A} \times e^{-\frac{1}{2}\left(\frac{\left|{\lambda - \lambda_0}\right|}{\sigma_{G}}\right)^2},
    \label{eq:g}
\end{equation}
where A is the Gaussian amplitude and $\lambda_0$ is the central wavelength of the emission line. $\sigma_{G}$ is the width parameter,
\begin{equation}
    \sigma_{G} = \frac{\mbox{FWHM}}{2\times(2\times\mbox{ln}(2))^{1/2}}.
    \label{eq:g_sig}
\end{equation}
We used a fixed FWHM and velocity dispersion in five different wavelength windows where a linear continuum was adopted. The LBT-LUCI line profiles generally have a long red tail, while the blue side of the emission line has a sharp cutoff. Despite this peculiar profile, we do not attempt to fit it, as the fluxes are measured well in comparison to direct integration. The main lines utilized in this wavelength range are \10830 and \pg, though others are presented. We show three examples of our final spectra in Figure \ref{fig:spectra}, where we measure the line fluxes in a spectrum where only LUCI1 was available, one where only LUCI2 was available, and one where we co-added data from LUCI1 and LUCI2. Note that UM461 has the
highest S/N near-IR spectrum. However, the targets that only used LUCI1 or LUCI2 still had minimal noise. We report our measured line ratios for 6 targets in Table \ref{tab:ratios}: 2 from LUCI1, 2 from LUCI2, and 2 from LUCI1+LUCI2. The entirety of the table, reporting all 48 line ratios, will be available in machine-readable format.

\begin{deluxetable*}{r r r r r r r r r}
\label{tab:ratios}
\small
\tabletypesize{\small}
\tablecaption{Measured line fluxes relative to P$\gamma$ and equivalent widths (EWs) for a subsample of our targets. The table is organized by LUCI1 targets, LUCI2 targets, and LUCI1+2 targets. The entirety of this table will be available in machine-readable format.}
\tablehead{
\CH{} & \CH{P$\delta$ 10050} & \CH{} & \CH{\10830} & \CH{} & \CH{P$\beta$ 12820} & \CH{}  & \CH{P$\gamma$ 10938} & \CH{} \\[-2ex]
\CH{ID} & \CH{Flux/P$\gamma$} & \CH{EW [\AA]} & \CH{Flux/P$\gamma$} & \CH{EW [\AA]} & \CH{Flux/P$\gamma$} & \CH{EW [\AA]} & \CH{Flux [$10^{-17}$ erg/s/cm$^2$]} &  \CH{EW [\AA]}}

\startdata
\multicolumn{9}{c}{LUCI1} \\
\tableline
Mrk 5         & 0.52$\pm$0.02     & 54.929  & 2.22$\pm$0.06     & 254.032  & 1.88$\pm$0.05     & 269.983 & 244.00$\pm$5.01     & 115.302 \\
UM133         & 0.57$\pm$0.02     & 99.673  & 2.63$\pm$0.08     & 517.914  & 1.73$\pm$0.05     & 414.887 & 104.00$\pm$2.35     & 200.438 \\
\tableline
\multicolumn{9}{c}{LUCI2} \\
\tableline
UM420         & 0.60$\pm$0.02     & 43.237  & 2.80$\pm$0.08     & 229.651  & 3.03$\pm$0.20     & 169.708 & 322.00$\pm$6.67     & 83.388  \\
HS 2236+1344        & 0.95$\pm$0.05     & 58.217  & 6.67$\pm$0.22     & 440.972  & 2.79$\pm$0.09     & 282.331 & 93.20$\pm$2.39      & 69.769  \\
\tableline
\multicolumn{9}{c}{LUCI1 + LUCI2} \\
\tableline
UGC4483       & 0.57$\pm$0.02     & 47.5    & 2.45$\pm$0.07     & 264.858  & 1.85$\pm$0.05     & 295.445 & 519.00$\pm$11.00    & 109.575 \\
SHOC357       & 0.57$\pm$0.02     & 78.381  & 8.47$\pm$0.24     & 1322.857 & 1.76$\pm$0.05     & 373.002 & 262.00$\pm$5.41     & 158.561 \\ \tableline
\enddata
\end{deluxetable*}

\section{Comparison to Literature}
\label{compare}

\citet{izot2013, izot2014} were the first to utilize \10830 in analyses of \yp. They use the TripleSpec spectrograph on the Apache Point Observatory (APO) telescope with a resolution of 3500, LUCI on the LBT with a resolution of 8460 in the \textit{J} band, and data from the European Southern Observatory (ESO) archives, including spectra from the VLT/Infrared Spectrometer and Array Camera (ISAAC) and the New Technology Telescope (NTT)/SOFI. Our configuration of LUCI uses the \textit{zJ} filter and achieves a resolving power of $\sim2100$. Both works follow similar reduction steps.

\begin{figure*}[t!]
\centering
\centerline{\includegraphics[width=\textwidth]{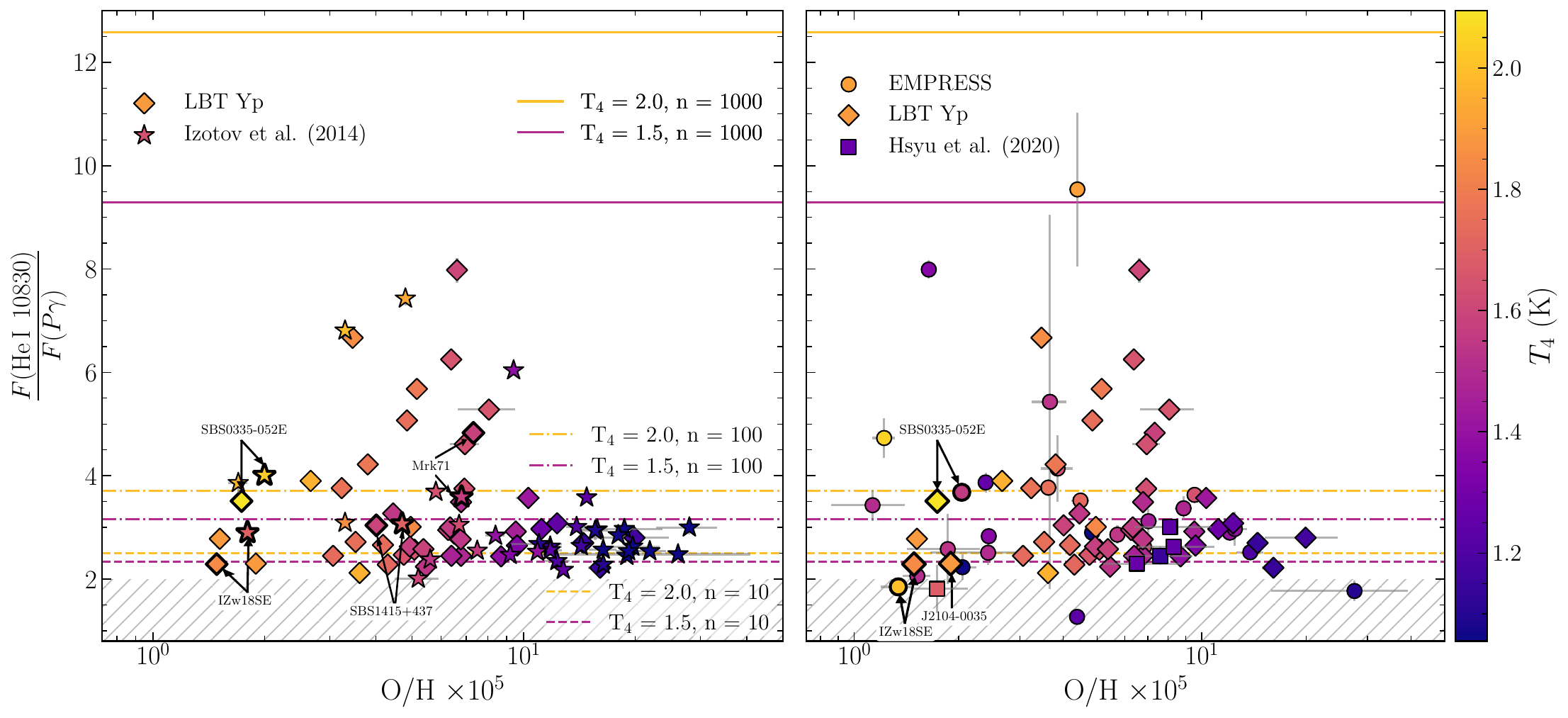}}
\caption{\textit{Left:} Flux ratio of \10830 to P$\gamma$ from this work as diamonds and \citet{izot2014} as stars. \textit{Right:} \citet{EMPRESS25} as circles compared to this work in diamonds. J2104-0035 has no counterpart in \citet{EMPRESS25} as they did not detect emission lines in their observations of this target. Targets from \citet{Hsyu2020} are shown as squares. Both panels are color-coded by temperature and shared observations between studies are bolded and labeled. The colored horizontal lines show constant temperatures and densities calculated with a nominal value of \y+ = 0.085 while the hatched region shows the nonphysical regime.}
\label{fig: ratio}
\end{figure*}

We compare in the left panel of Figure \ref{fig: ratio} the line ratios of \10830 to P$\gamma$ as a function of O/H for \citet{izot2014} as stars and this work as diamonds. We color-code by the reported $T_e$ and highlight those targets shared among the samples. It is evident our sample is more populated at lower metallicities; however, the median uncertainty in this work is $\tilde{\sigma} = 0.08$, which is comparable to those presented in \citet{izot2014} who have a median uncertainty of $\tilde{\sigma} = 0.05$.

\citet{Hsyu2020} also gather NIR spectra to measure the \10830 line. They observe 16 galaxies in low-resolution mode using Keck to complement their 32 optical spectra. They also reduce the near-IR spectra using {\tt PypeIt}, though they do not mention telluric correction. They combine their Keck results with Sloan Digital Sky Survey (SDSS) data to achieve a larger sample size. Out of their two samples (e.g., their Sample 1 and Sample 2), only five are from their Keck sample with NIR counterparts. We do not share any of these targets among our sample, but plot them in the right panel of Figure \ref{fig: ratio} as squares. 

More recently, the EMPRESS program \citep{EMPRESS25} gathers observations of \10830 in 29 targets using Subaru (with a maximum spectral resolution $\lesssim$ 1200), though emission lines were not detected in 2. In the right panel of Figure \ref{fig: ratio}, we compare line ratios of \10830 to P$\gamma$, this work shown as diamonds and \citet{EMPRESS25} shown as circles, again color-coding by $T_e$. The median error in \citet{EMPRESS25} is $\tilde{\sigma} = 0.18$, which is more than twice the value of this work, likely due to their low resolution, but they also do not mention telluric correction. Further, we target 3 of the same galaxies, those of which are labeled. Note that our J2104-0335 point has no circular counterpart as this was one of the targets that \citet{EMPRESS25} did not detect emission lines. Finally, in \citet{EMPRESS25}, only 10 of their 29 galaxies are included in their final sample, whereas over two-thirds of our targets qualify \citep[][Paper IV]{Aver2026}. \citet{EMPRESS25} combines their sample with that of \citet{Hsyu2020} to complete their measurement of \yp. See Appendix A in Paper~I \citep{Skillman2026} for a more complete description of the differences between our work, \citet{Hsyu2020}, and \citet{EMPRESS25}.

The horizontal lines in both panels show regions of constant temperatures and densities. These were calculated with emissivity data from \citet{zanna22} and a chosen value of $y^+ = 0.085$ to predict a flux ratio given $\frac{F(10830)}{F(P\gamma)} \propto y^+ \frac{E(10830)}{E(P\gamma)}$. The lines are color-coded by temperature but have different styles based on density. The hatched region in both panels represents a nonphysical regime of possible line ratios because of emission line physics. The emissivity of \10830 is larger than that of P$\gamma$ because \10830 arises from a metastable level, leading to a surplus of electrons ready to be excited. Therefore, despite helium being lower in abundance than hydrogen, the \10830 line is stronger. For realistic physical conditions, this leads to a floor around $\sim2$. Three \citet{EMPRESS25} targets and one \citet{Hsyu2020} target fall into this nonphysical regime; however, none of our targets or the objects from \citet{izot2013, izot2014} fall into this category. This once again illustrates the strength of our sample over previous studies.

\section{Conclusions}
\label{conclusion}
In this paper, we describe the LBT-LUCI observations of 48 low-metallicity galaxies observed as part of the LBT \yp\ Project. These galaxies were observed to measure the flux ratio of \10830 to the nearby \pg\ line due to the sensitivity of \10830 to density. Our key steps and findings are:
\begin{enumerate}
    \item We use {\tt PypeIt} to homogeneously reduce our near-IR LUCI spectra. We wavelength calibrate by-hand, correct for telluric absorption caused by Earth's atmosphere, and co-add the spectra when both LUCI1 and LUCI2 were available. Finally, we use a Gaussian profile to measure the line fluxes. 
    \item We compare the results of our line ratios to previous literature, namely \citet{izot2014}, \citet{Hsyu2020}, and the recent \citet{EMPRESS25} study. Our results are fully consistent with the expected trends from atomic data, while some points from the aforementioned studies reside in a non-physical regime.
\end{enumerate}
Overall, our efforts have substantially expanded the available measurements of \10830 to \pg\ line ratios in near-IR, low-metallicity spectra.

\begin{acknowledgements}
This work was supported by funds provided by NSF Collaborative Research Grants AST-2205817 to RWP, AST-2205864 to EDS, and AST-2205958 to EA.
Our team workshop at OSU in July 2024 was sponsored in part by OSU's Center for Cosmology and AstroParticle Physics (CCAPP). We are grateful to Olga Kuhn at LBTO who helped us navigate the intricacies of creating LUCI observing scripts.

This work is based on observations made with the Large Binocular Telescope. The LBT is an international collaboration among institutions in the United States, Italy and Germany. LBT Corporation Members are: The University of Arizona on behalf of the Arizona Board of Regents; Istituto Nazionale di Astrofisica, Italy; LBT Beteiligungsgesellschaft, Germany, representing the Max-Planck Society, The Leibniz Institute for Astrophysics Potsdam, and Heidelberg University; The Ohio State University, and The Research Corporation, on behalf of The University of Notre Dame, University of Minnesota and University of Virginia. Observations have benefited from the use of ALTA Center (alta.arcetri.inaf.it) forecasts performed with the Astro-Meso-Nh model. Initialization data of the ALTA automatic forecast system come from the General Circulation Model (HRES) of the European Centre for Medium Range Weather Forecasts.

This research used the facilities of the Italian Center for Astronomical Archive (IA2) operated by INAF at the Astronomical Observatory of Trieste.

EDS, EA, DAB, NSJR, and JHM would like to acknowledge and thank Stanley Hubbard for his generous gift to
the University of Minnesota that allowed the University to become a member of the LBT collaboration.

This research made use of {\ttfamily PypeIt},\footnote{\url{https://pypeit.readthedocs.io/en/latest/}}
a Python package for semi-automated reduction of astronomical slit-based spectroscopy
\citep{pypeit:joss_pub, pypeit:zenodo}.

\end{acknowledgements}

\facilities{LBT (MODS), LBT (LUCI)}
\software{
\texttt{astropy} \citep{astr2013, astr2018, astr2022},
\texttt{jupyter} \citep{kluy2016},
\texttt{modsIDL} \citep{crox2019},
\texttt{modsCCDRed} \citep{pogg2019},
\texttt{PypeIt} \citep{pypeit:joss_arXiv, pypeit:joss_pub, pypeit:zenodo},
\texttt{PyNeb} \citep{luri2012,luri2015L},
\texttt{numpy} \citep{harr2020}
}


\bibliographystyle{aasjournalv7}
\bibliography{yp_total_bib}

@ARTICLE{astr2013,
       author = {{Astropy Collaboration} and {Robitaille}, Thomas P. and {Tollerud}, Erik J. and {Greenfield}, Perry and {Droettboom}, Michael and {Bray}, Erik and {Aldcroft}, Tom and {Davis}, Matt and {Ginsburg}, Adam and {Price-Whelan}, Adrian M. and {Kerzendorf}, Wolfgang E. and {Conley}, Alexander and {Crighton}, Neil and {Barbary}, Kyle and {Muna}, Demitri and {Ferguson}, Henry and {Grollier}, Fr{\'e}d{\'e}ric and {Parikh}, Madhura M. and {Nair}, Prasanth H. and {Unther}, Hans M. and {Deil}, Christoph and {Woillez}, Julien and {Conseil}, Simon and {Kramer}, Roban and {Turner}, James E.~H. and {Singer}, Leo and {Fox}, Ryan and {Weaver}, Benjamin A. and {Zabalza}, Victor and {Edwards}, Zachary I. and {Azalee Bostroem}, K. and {Burke}, D.~J. and {Casey}, Andrew R. and {Crawford}, Steven M. and {Dencheva}, Nadia and {Ely}, Justin and {Jenness}, Tim and {Labrie}, Kathleen and {Lim}, Pey Lian and {Pierfederici}, Francesco and {Pontzen}, Andrew and {Ptak}, Andy and {Refsdal}, Brian and {Servillat}, Mathieu and {Streicher}, Ole},
        title = "{Astropy: A community Python package for astronomy}",
      journal = {\aap},
         year = 2013,
        month = oct,
       volume = {558},
          eid = {A33},
        pages = {A33},
          doi = {10.1051/0004-6361/201322068},
archivePrefix = {arXiv},
       eprint = {1307.6212},
 primaryClass = {astro-ph.IM},
       adsurl = {https://ui.adsabs.harvard.edu/abs/2013A&A...558A..33A}
}

@ARTICLE{astr2018,
       author = {{Astropy Collaboration} and {Price-Whelan}, A.~M. and {Sip{\H{o}}cz}, B.~M. and {G{\"u}nther}, H.~M. and {Lim}, P.~L. and {Crawford}, S.~M. and {Conseil}, S. and {Shupe}, D.~L. and {Craig}, M.~W. and {Dencheva}, N. and {Ginsburg}, A. and {VanderPlas}, J.~T. and {Bradley}, L.~D. and {P{\'e}rez-Su{\'a}rez}, D. and {de Val-Borro}, M. and {Aldcroft}, T.~L. and {Cruz}, K.~L. and {Robitaille}, T.~P. and {Tollerud}, E.~J. and {Ardelean}, C. and {Babej}, T. and {Bach}, Y.~P. and {Bachetti}, M. and {Bakanov}, A.~V. and {Bamford}, S.~P. and {Barentsen}, G. and {Barmby}, P. and {Baumbach}, A. and {Berry}, K.~L. and {Biscani}, F. and {Boquien}, M. and {Bostroem}, K.~A. and {Bouma}, L.~G. and {Brammer}, G.~B. and {Bray}, E.~M. and {Breytenbach}, H. and {Buddelmeijer}, H. and {Burke}, D.~J. and {Calderone}, G. and {Cano Rodr{\'\i}guez}, J.~L. and {Cara}, M. and {Cardoso}, J.~V.~M. and {Cheedella}, S. and {Copin}, Y. and {Corrales}, L. and {Crichton}, D. and {D'Avella}, D. and {Deil}, C. and {Depagne}, {\'E}. and {Dietrich}, J.~P. and {Donath}, A. and {Droettboom}, M. and {Earl}, N. and {Erben}, T. and {Fabbro}, S. and {Ferreira}, L.~A. and {Finethy}, T. and {Fox}, R.~T. and {Garrison}, L.~H. and {Gibbons}, S.~L.~J. and {Goldstein}, D.~A. and {Gommers}, R. and {Greco}, J.~P. and {Greenfield}, P. and {Groener}, A.~M. and {Grollier}, F. and {Hagen}, A. and {Hirst}, P. and {Homeier}, D. and {Horton}, A.~J. and {Hosseinzadeh}, G. and {Hu}, L. and {Hunkeler}, J.~S. and {Ivezi{\'c}}, {\v{Z}}. and {Jain}, A. and {Jenness}, T. and {Kanarek}, G. and {Kendrew}, S. and {Kern}, N.~S. and {Kerzendorf}, W.~E. and {Khvalko}, A. and {King}, J. and {Kirkby}, D. and {Kulkarni}, A.~M. and {Kumar}, A. and {Lee}, A. and {Lenz}, D. and {Littlefair}, S.~P. and {Ma}, Z. and {Macleod}, D.~M. and {Mastropietro}, M. and {McCully}, C. and {Montagnac}, S. and {Morris}, B.~M. and {Mueller}, M. and {Mumford}, S.~J. and {Muna}, D. and {Murphy}, N.~A. and {Nelson}, S. and {Nguyen}, G.~H. and {Ninan}, J.~P. and {N{\"o}the}, M. and {Ogaz}, S. and {Oh}, S. and {Parejko}, J.~K. and {Parley}, N. and {Pascual}, S. and {Patil}, R. and {Patil}, A.~A. and {Plunkett}, A.~L. and {Prochaska}, J.~X. and {Rastogi}, T. and {Reddy Janga}, V. and {Sabater}, J. and {Sakurikar}, P. and {Seifert}, M. and {Sherbert}, L.~E. and {Sherwood-Taylor}, H. and {Shih}, A.~Y. and {Sick}, J. and {Silbiger}, M.~T. and {Singanamalla}, S. and {Singer}, L.~P. and {Sladen}, P.~H. and {Sooley}, K.~A. and {Sornarajah}, S. and {Streicher}, O. and {Teuben}, P. and {Thomas}, S.~W. and {Tremblay}, G.~R. and {Turner}, J.~E.~H. and {Terr{\'o}n}, V. and {van Kerkwijk}, M.~H. and {de la Vega}, A. and {Watkins}, L.~L. and {Weaver}, B.~A. and {Whitmore}, J.~B. and {Woillez}, J. and {Zabalza}, V. and {Astropy Contributors}},
        title = "{The Astropy Project: Building an Open-science Project and Status of the v2.0 Core Package}",
      journal = {\aj},
         year = 2018,
        month = sep,
       volume = {156},
       number = {3},
          eid = {123},
        pages = {123},
          doi = {10.3847/1538-3881/aabc4f},
archivePrefix = {arXiv},
       eprint = {1801.02634},
 primaryClass = {astro-ph.IM},
       adsurl = {https://ui.adsabs.harvard.edu/abs/2018AJ....156..123A}
}

@ARTICLE{astr2022,
       author = {{Astropy Collaboration} and {Price-Whelan}, Adrian M. and {Lim}, Pey Lian and {Earl}, Nicholas and {Starkman}, Nathaniel and {Bradley}, Larry and {Shupe}, David L. and {Patil}, Aarya A. and {Corrales}, Lia and {Brasseur}, C.~E. and {N{\"o}the}, Maximilian and {Donath}, Axel and {Tollerud}, Erik and {Morris}, Brett M. and {Ginsburg}, Adam and {Vaher}, Eero and {Weaver}, Benjamin A. and {Tocknell}, James and {Jamieson}, William and {van Kerkwijk}, Marten H. and {Robitaille}, Thomas P. and {Merry}, Bruce and {Bachetti}, Matteo and {G{\"u}nther}, H. Moritz and {Aldcroft}, Thomas L. and {Alvarado-Montes}, Jaime A. and {Archibald}, Anne M. and {B{\'o}di}, Attila and {Bapat}, Shreyas and {Barentsen}, Geert and {Baz{\'a}n}, Juanjo and {Biswas}, Manish and {Boquien}, M{\'e}d{\'e}ric and {Burke}, D.~J. and {Cara}, Daria and {Cara}, Mihai and {Conroy}, Kyle E. and {Conseil}, Simon and {Craig}, Matthew W. and {Cross}, Robert M. and {Cruz}, Kelle L. and {D'Eugenio}, Francesco and {Dencheva}, Nadia and {Devillepoix}, Hadrien A.~R. and {Dietrich}, J{\"o}rg P. and {Eigenbrot}, Arthur Davis and {Erben}, Thomas and {Ferreira}, Leonardo and {Foreman-Mackey}, Daniel and {Fox}, Ryan and {Freij}, Nabil and {Garg}, Suyog and {Geda}, Robel and {Glattly}, Lauren and {Gondhalekar}, Yash and {Gordon}, Karl D. and {Grant}, David and {Greenfield}, Perry and {Groener}, Austen M. and {Guest}, Steve and {Gurovich}, Sebastian and {Handberg}, Rasmus and {Hart}, Akeem and {Hatfield-Dodds}, Zac and {Homeier}, Derek and {Hosseinzadeh}, Griffin and {Jenness}, Tim and {Jones}, Craig K. and {Joseph}, Prajwel and {Kalmbach}, J. Bryce and {Karamehmetoglu}, Emir and {Ka{\l}uszy{\'n}ski}, Miko{\l}aj and {Kelley}, Michael S.~P. and {Kern}, Nicholas and {Kerzendorf}, Wolfgang E. and {Koch}, Eric W. and {Kulumani}, Shankar and {Lee}, Antony and {Ly}, Chun and {Ma}, Zhiyuan and {MacBride}, Conor and {Maljaars}, Jakob M. and {Muna}, Demitri and {Murphy}, N.~A. and {Norman}, Henrik and {O'Steen}, Richard and {Oman}, Kyle A. and {Pacifici}, Camilla and {Pascual}, Sergio and {Pascual-Granado}, J. and {Patil}, Rohit R. and {Perren}, Gabriel I. and {Pickering}, Timothy E. and {Rastogi}, Tanuj and {Roulston}, Benjamin R. and {Ryan}, Daniel F. and {Rykoff}, Eli S. and {Sabater}, Jose and {Sakurikar}, Parikshit and {Salgado}, Jes{\'u}s and {Sanghi}, Aniket and {Saunders}, Nicholas and {Savchenko}, Volodymyr and {Schwardt}, Ludwig and {Seifert-Eckert}, Michael and {Shih}, Albert Y. and {Jain}, Anany Shrey and {Shukla}, Gyanendra and {Sick}, Jonathan and {Simpson}, Chris and {Singanamalla}, Sudheesh and {Singer}, Leo P. and {Singhal}, Jaladh and {Sinha}, Manodeep and {Sip{\H{o}}cz}, Brigitta M. and {Spitler}, Lee R. and {Stansby}, David and {Streicher}, Ole and {{\v{S}}umak}, Jani and {Swinbank}, John D. and {Taranu}, Dan S. and {Tewary}, Nikita and {Tremblay}, Grant R. and {de Val-Borro}, Miguel and {Van Kooten}, Samuel J. and {Vasovi{\'c}}, Zlatan and {Verma}, Shresth and {de Miranda Cardoso}, Jos{\'e} Vin{\'\i}cius and {Williams}, Peter K.~G. and {Wilson}, Tom J. and {Winkel}, Benjamin and {Wood-Vasey}, W.~M. and {Xue}, Rui and {Yoachim}, Peter and {Zhang}, Chen and {Zonca}, Andrea and {Astropy Project Contributors}},
        title = "{The Astropy Project: Sustaining and Growing a Community-oriented Open-source Project and the Latest Major Release (v5.0) of the Core Package}",
      journal = {\apj},
         year = 2022,
        month = aug,
       volume = {935},
       number = {2},
          eid = {167},
        pages = {167},
          doi = {10.3847/1538-4357/ac7c74},
archivePrefix = {arXiv},
       eprint = {2206.14220},
 primaryClass = {astro-ph.IM},
       adsurl = {https://ui.adsabs.harvard.edu/abs/2022ApJ...935..167A}
}

@ARTICLE{aver2010,
       author = {{Aver}, Erik and {Olive}, Keith A. and {Skillman}, Evan D.},
        title = "{A new approach to systematic uncertainties and self-consistency in helium abundance determinations}",
      journal = {\jcap},
         year = 2010,
        month = may,
       volume = {2010},
       number = {5},
          eid = {003},
        pages = {003},
          doi = {10.1088/1475-7516/2010/05/003},
archivePrefix = {arXiv},
       eprint = {1001.5218},
 primaryClass = {astro-ph.CO},
       adsurl = {https://ui.adsabs.harvard.edu/abs/2010JCAP...05..003A}
}

@ARTICLE{aver2011,
       author = {{Aver}, Erik and {Olive}, Keith A. and {Skillman}, Evan D.},
        title = "{Mapping systematic errors in helium abundance determinations using Markov Chain Monte Carlo}",
      journal = {\jcap},
         year = 2011,
        month = mar,
       volume = {2011},
       number = {3},
          eid = {043},
        pages = {043},
          doi = {10.1088/1475-7516/2011/03/043},
archivePrefix = {arXiv},
       eprint = {1012.2385},
 primaryClass = {astro-ph.CO},
       adsurl = {https://ui.adsabs.harvard.edu/abs/2011JCAP...03..043A}
}

@ARTICLE{aver2021,
       author = {{Aver}, Erik and {Berg}, Danielle A. and {Olive}, Keith A. and {Pogge}, Richard W. and {Salzer}, John J. and {Skillman}, Evan D.},
        title = "{Improving helium abundance determinations with Leo P as a case study}",
      journal = {\jcap},
         year = 2021,
        month = mar,
       volume = {2021},
       number = {3},
          eid = {027},
        pages = {027},
          doi = {10.1088/1475-7516/2021/03/027},
archivePrefix = {arXiv},
       eprint = {2010.04180},
 primaryClass = {astro-ph.CO},
       adsurl = {https://ui.adsabs.harvard.edu/abs/2021JCAP...03..027A}
}

@ARTICLE{aver2022,
       author = {{Aver}, Erik and {Berg}, Danielle A. and {Hirschauer}, Alec S. and {Olive}, Keith A. and {Pogge}, Richard W. and {Rogers}, Noah S.~J. and {Salzer}, John J. and {Skillman}, Evan D.},
        title = "{A comprehensive chemical abundance analysis of the extremely metal poor Leoncino Dwarf galaxy (AGC 198691)}",
      journal = {\mnras},
         year = 2022,
        month = feb,
       volume = {510},
       number = {1},
        pages = {373-382},
          doi = {10.1093/mnras/stab3226},
archivePrefix = {arXiv},
       eprint = {2109.00178},
 primaryClass = {astro-ph.GA},
       adsurl = {https://ui.adsabs.harvard.edu/abs/2022MNRAS.510..373A}
}

@ARTICLE{aver2015,
       author = {{Aver}, Erik and {Olive}, Keith A. and {Skillman}, Evan D.},
        title = "{The effects of He I {\ensuremath{\lambda}}10830 on helium abundance determinations}",
      journal = {\jcap},
         year = 2015,
        month = jul,
       volume = {2015},
       number = {7},
        pages = {011-011},
          doi = {10.1088/1475-7516/2015/07/011},
archivePrefix = {arXiv},
       eprint = {1503.08146},
 primaryClass = {astro-ph.CO},
       adsurl = {https://ui.adsabs.harvard.edu/abs/2015JCAP...07..011A}
}

@ARTICLE{vanD2001,
       author = {{van Dokkum}, Pieter G.},
        title = "{Cosmic-Ray Rejection by Laplacian Edge Detection}",
      journal = {\pasp},
         year = 2001,
        month = nov,
       volume = {113},
       number = {789},
        pages = {1420-1427},
          doi = {10.1086/323894},
archivePrefix = {arXiv},
       eprint = {astro-ph/0108003},
 primaryClass = {astro-ph},
       adsurl = {https://ui.adsabs.harvard.edu/abs/2001PASP..113.1420V}
}

@ARTICLE{EMPRESS25,
       author = {{Yanagisawa}, Hiroto and {Ouchi}, Masami and {Matsumoto}, Akinori and {Kawasaki}, Masahiro and {Murai}, Kai and {Nakajima}, Kimihiko and {Kohri}, Kazunori and {Sugahara}, Yuma and {Nagamine}, Kentaro and {Tanaka}, Ichi and {Kim}, Ji Hoon and {Ono}, Yoshiaki and {Nakane}, Minami and {Fukushima}, Keita and {Harikane}, Yuichi and {Hirai}, Yutaka and {Isobe}, Yuki and {Kusakabe}, Haruka and {Onodera}, Masato and {Rauch}, Michael and {Yajima}, Hidenobu},
        title = "{EMPRESS. XV. A New Determination of the Primordial Helium Abundance Suggesting a Moderately Low $Y_\mathrm{P}$ Value}",
      journal = {arXiv e-prints},
         year = 2025,
        month = jun,
          eid = {arXiv:2506.24050},
        pages = {arXiv:2506.24050},
          doi = {10.48550/arXiv.2506.24050},
archivePrefix = {arXiv},
       eprint = {2506.24050},
 primaryClass = {astro-ph.GA},
       adsurl = {https://ui.adsabs.harvard.edu/abs/2025arXiv250624050Y}
}

@software{crox2019,
  author       = {Kevin V. Croxall and
                  Richard W. Pogge},
  title        = {rwpogge/modsIDL: modsIDL Binocular Release},
  month        = feb,
  year         = 2019,
  publisher    = {Zenodo},
  version      = {v1.0},
  doi          = {10.5281/zenodo.2561424},
  url          = {https://doi.org/10.5281/zenodo.2561424}
}

@ARTICLE{harr2020,
       author = {{Harris}, Charles R. and {Millman}, K. Jarrod and {van der Walt}, St{\'e}fan J. and {Gommers}, Ralf and {Virtanen}, Pauli and {Cournapeau}, David and {Wieser}, Eric and {Taylor}, Julian and {Berg}, Sebastian and {Smith}, Nathaniel J. and {Kern}, Robert and {Picus}, Matti and {Hoyer}, Stephan and {van Kerkwijk}, Marten H. and {Brett}, Matthew and {Haldane}, Allan and {del R{\'\i}o}, Jaime Fern{\'a}ndez and {Wiebe}, Mark and {Peterson}, Pearu and {G{\'e}rard-Marchant}, Pierre and {Sheppard}, Kevin and {Reddy}, Tyler and {Weckesser}, Warren and {Abbasi}, Hameer and {Gohlke}, Christoph and {Oliphant}, Travis E.},
        title = "{Array programming with NumPy}",
      journal = {\nat},
         year = 2020,
        month = sep,
       volume = {585},
       number = {7825},
        pages = {357-362},
          doi = {10.1038/s41586-020-2649-2},
archivePrefix = {arXiv},
       eprint = {2006.10256},
 primaryClass = {cs.MS},
       adsurl = {https://ui.adsabs.harvard.edu/abs/2020Natur.585..357H}
}

@ARTICLE{izot2014,
       author = {{Izotov}, Y.~I. and {Thuan}, T.~X. and {Guseva}, N.~G.},
        title = "{A new determination of the primordial He abundance using the He I {\ensuremath{\lambda}}10830 {\r{A}} emission line: cosmological implications}",
      journal = {\mnras},
         year = 2014,
        month = nov,
       volume = {445},
       number = {1},
        pages = {778-793},
          doi = {10.1093/mnras/stu1771},
archivePrefix = {arXiv},
       eprint = {1408.6953},
 primaryClass = {astro-ph.CO},
       adsurl = {https://ui.adsabs.harvard.edu/abs/2014MNRAS.445..778I}
}

@conference{kluy2016, 
Title = {Jupyter Notebooks -- a publishing format for reproducible computational workflows}, 
Author = {Thomas Kluyver and Benjamin Ragan-Kelley and Fernando P{\'e}rez and Brian Granger and Matthias Bussonnier and Jonathan Frederic and Kyle Kelley and Jessica Hamrick and Jason Grout and Sylvain Corlay and Paul Ivanov and Dami{\'a}n Avila and Safia Abdalla and Carol Willing}, 
Booktitle = {Positioning and Power in Academic Publishing: Players, Agents and Agendas}, 
Editor = {F. Loizides and B. Schmidt}, 
Organization = {IOS Press}, 
Pages = {87 - 90}, 
Year = {2016} }

@ARTICLE{luri2012,
       author = {{Luridiana}, Valentina and {Morisset}, Christophe and {Shaw}, Richard A.},
        title = "{PyNeb: a new software for the analysis of emission lines}",
      journal = {IAU Symposium},
         year = 2012,
        month = aug,
       volume = {283},
        pages = {422-423},
          doi = {10.1017/S1743921312011738},
       adsurl = {https://ui.adsabs.harvard.edu/abs/2012IAUS..283..422L}
}

@ARTICLE{luri2015L,
       author = {{Luridiana}, V. and {Morisset}, C. and {Shaw}, R.~A.},
        title = "{PyNeb: a new tool for analyzing emission lines. I. Code description and validation of results}",
      journal = {\aap},
         year = 2015,
        month = jan,
       volume = {573},
          eid = {A42},
        pages = {A42},
          doi = {10.1051/0004-6361/201323152},
archivePrefix = {arXiv},
       eprint = {1410.6662},
 primaryClass = {astro-ph.IM},
       adsurl = {https://ui.adsabs.harvard.edu/abs/2015A&A...573A..42L}
}

@software{pogg2019,
       author = {{Pogge}, Richard},
        title = "{rwpogge/modsCCDRed: v2.0.1}",
         year = 2019,
        month = apr,
          eid = {10.5281/zenodo.2647501},
          doi = {10.5281/zenodo.2647501},
      version = {2.0.1},
    publisher = {Zenodo},
       adsurl = {https://ui.adsabs.harvard.edu/abs/2019zndo...2647501P}
}

@ARTICLE{izot2013,
       author = {{Izotov}, Y.~I. and {Stasi{\'n}ska}, G. and {Guseva}, N.~G.},
        title = "{Primordial $^{4}$He abundance: a determination based on the largest sample of H II regions with a methodology tested on model H II regions}",
      journal = {\aap},
         year = 2013,
        month = oct,
       volume = {558},
          eid = {A57},
        pages = {A57},
          doi = {10.1051/0004-6361/201220782},
archivePrefix = {arXiv},
       eprint = {1308.2100},
 primaryClass = {astro-ph.CO},
       adsurl = {https://ui.adsabs.harvard.edu/abs/2013A&A...558A..57I}
}

@ARTICLE{Hsyu2020,
       author = {{Hsyu}, Tiffany and {Cooke}, Ryan J. and {Prochaska}, J. Xavier and {Bolte}, Michael},
        title = "{The PHLEK Survey: A New Determination of the Primordial Helium Abundance}",
      journal = {\apj},
         year = 2020,
        month = jun,
       volume = {896},
       number = {1},
          eid = {77},
        pages = {77},
          doi = {10.3847/1538-4357/ab91af},
archivePrefix = {arXiv},
       eprint = {2005.12290},
 primaryClass = {astro-ph.GA},
       adsurl = {https://ui.adsabs.harvard.edu/abs/2020ApJ...896...77H}
}

@ARTICLE{pypeit:joss_arXiv,
       author = {{Prochaska}, J. Xavier and {Hennawi}, Joseph F. and {Westfall}, Kyle B. and
         {Cooke}, Ryan J. and {Wang}, Feige and {Hsyu}, Tiffany and
         {Davies}, Frederick B. and {Farina}, Emanuele Paolo},
       shortauthor = {Prochaska et al.},
        title = "{PypeIt: The Python Spectroscopic Data Reduction Pipeline}",
      journal = {arXiv e-prints},
         year = {2020a},
        month = may,
          eid = {arXiv:2005.06505},
        pages = {arXiv:2005.06505},
archivePrefix = {arXiv},
       eprint = {2005.06505},
 primaryClass = {astro-ph.IM},
       adsurl = {https://ui.adsabs.harvard.edu/abs/2020arXiv200506505P}
}

@article{pypeit:joss_pub,
    doi = {10.21105/joss.02308},
    url = {https://doi.org/10.21105/joss.02308},
    year =  {2020c},
    month = dec,
    publisher = {The Open Journal},
    volume = {5},
    number = {56},
    pages = {2308},
    author = {J. Xavier Prochaska and Joseph F. Hennawi and Kyle B. Westfall and Ryan J. Cooke and Feige Wang and Tiffany Hsyu and Frederick B. Davies and Emanuele Paolo Farina and Debora Pelliccia},
    shortauthor = {Prochaska et al.},
    title = {PypeIt: The Python Spectroscopic Data Reduction Pipeline},
    journal = {Journal of Open Source Software}
}

@MISC{pypeit:zenodo,
       author = {{Prochaska}, J. Xavier and {Hennawi}, Joseph and {Cooke}, Ryan and
         {Westfall}, Kyle and {Wang}, Feige and {EmAstro} and {Tiffanyhsyu} and
         {Wasserman}, Asher and {Villaume}, Alexa and {Marijana777} and
         {Schindler}, JT and {Young}, David and {Simha}, Sunil and
         {Wilde}, Matt and {Tejos}, Nicolas and {Isbell}, Jacob and
         {Fl{\"o}rs}, Andreas and {Sandford}, Nathan and {Vasovi{\'c}}, Zlatan and
         {Betts}, Edward and {Holden}, Brad},
        shortauthor = {Prochaska et al.},
        title = "{pypeit/PypeIt: Release 1.0.0}",
         year = {2020b},
        month = apr,
          eid = {10.5281/zenodo.3743493},
          doi = {10.5281/zenodo.3743493},
      version = {v1.0.0},
    publisher = {Zenodo},
       adsurl = {https://ui.adsabs.harvard.edu/abs/2020zndo...3743493P}
}

@ARTICLE{Peimbert1974,
       author = {{Peimbert}, M. and {Torres-Peimbert}, S.},
        title = "{Chemical composition of H II regions in the Large Magellanic Cloud and its cosmological implications.}",
      journal = {\apj},
         year = 1974,
        month = oct,
       volume = {193},
        pages = {327-333},
          doi = {10.1086/153166},
       adsurl = {https://ui.adsabs.harvard.edu/abs/1974ApJ...193..327P}
}

@ARTICLE{Peimbert1976,
       author = {{Peimbert}, M. and {Torres-Peimbert}, S.},
        title = "{Chemical composition of H II regions in the Small Magellanic Cloud and the pregalactic helium abundance.}",
      journal = {\apj},
         year = 1976,
        month = feb,
       volume = {203},
        pages = {581-586},
          doi = {10.1086/154114},
       adsurl = {https://ui.adsabs.harvard.edu/abs/1976ApJ...203..581P}
}

@ARTICLE{zanna22,
       author = {{Del Zanna}, G. and {Storey}, P.~J.},
        title = "{Helium line emissivities for nebular astrophysics}",
      journal = {\mnras},
         year = 2022,
        month = jun,
       volume = {513},
       number = {1},
        pages = {1198-1209},
          doi = {10.1093/mnras/stac800},
archivePrefix = {arXiv},
       eprint = {2204.01537},
 primaryClass = {astro-ph.GA},
       adsurl = {https://ui.adsabs.harvard.edu/abs/2022MNRAS.513.1198D}
}

@article{Skillman2026,
    author = "Skillman, Evan and others",
    title = "{The LBT $Y_{\rm p}$ Project I: An Improved Determinaton of the Primordial Helium Abundance -- Project Description, Sample Selection, Observations, and Methodology}",
    year = "2026"
}

@article{Aver2026,
    author = "Aver, Erik and others",
    title = "{The LBT $Y_{\rm p}$ Project IV: A New Value of the Primordial Helium Abundance}",
    year = "2026"
}

\clearpage

\end{document}